\documentclass[
draft,
]{agujournal2025}
\usepackage[normalem]{ulem}

\begin{document}

\journalname{AGU Advances}

%%%%%%%%%%%%%%%%%%%%%%%%%%%%%%%%%%%%%%%%%%%%%%%
%  TITLE
%%%%%%%%%%%%%%%%%%%%%%%%%%%%%%%%%%%%%%%%%%%%%%%

% Embedding Tacit Expertise in Auditable Workflows: A Skill-Orchestrated HydroAgent for Flood Forecasting
\title{Community HydroAgent: Formalizing Forecaster Expertise into Skill-Orchestrated Flood Forecasting Workflows}

%%%%%%%%%%%%%%%%%%%%%%%%%%%%%%%%%%%%%%%%%%%%%%%
%  AUTHORS AND AFFILIATIONS
%%%%%%%%%%%%%%%%%%%%%%%%%%%%%%%%%%%%%%%%%%%%%%%

% List authors by first name or initial followed by last name and
% separated by commas. Use \affil{} to number affiliations, and
% \thanks{} for author notes.
%
% Example: \authors{A. B. Author\affil{1}\thanks{Current address, Antarctica},
%                   B. C. Author\affil{2,3},
%                   and D. E. Author\affil{3,4}\thanks{Also funded by Monsanto.}}

\authors{Qingyi Yang\affil{1},
         Siqian Qiu\affil{2},
         Bing Li\affil{3},
         Xu Shan\affil{4},
         Jia Feng\affil{5},
         Shunan Zhou\affil{6,7},
         Xudong Zhou\affil{8},
         Tiantian Xing\affil{1},
         Jiale Guo\affil{1},
         Xiaoyi Dong\affil{9},
         Gaoyu Liu\affil{10},
         Xiaohuan Liu\affil{1},
         Haiqing Pu\affil{5},
         Qingwen Deng\affil{11},
         Xun Zhang\affil{12,13},
         Zhongrun Xiang\affil{14},
         Haiyang Qian\affil{15,16},
         Ying Yan\affil{17},
         Yongkang Xu\affil{18},
         Nuo Lei\affil{13},
         Tianlong Jia\affil{19},
         Baoying Shan\affil{1},
         and Carlo De Michele\affil{1}}

% \affiliation{1}{First Affiliation}
% \affiliation{2}{Second Affiliation}

\affiliation{1}{Department of Civil and Environmental Engineering (D.I.C.A.), Politecnico di Milano, 20133 Milano MI, Italy}
\affiliation{2}{College of Hydraulic and Environmental Engineering, China Three Gorges University, 443002 Yichang, China}
\affiliation{3}{Independent Researcher, Shandong, 252300, China}
\affiliation{4}{Delft University of Technology, Delft, the Netherlands}
\affiliation{5}{Qinhuangdao Hydrological Survey and Research Center of Hebei Province, Qinhuangdao, China}
\affiliation{6}{School of Hydraulic Engineering, Dalian University of Technology, 116024 Dalian, Liaoning, China}
\affiliation{7}{Institute of Photogrammetry and Remote Sensing, TU Dresden University of Technology, 01062 Dresden, Germany}
\affiliation{8}{Institute of Hydraulic and Ocean Engineering, Ningbo University, Ningbo 315211, China}
\affiliation{9}{China IPPR International Engineering Co., Ltd., SINOMACH, Beijing, China}
\affiliation{10}{College of Hydrology and Water Resources, Hohai University, Nanjing, China}
\affiliation{11}{School of Earth Science and Engineering, Nanjing University, Nanjing, China}
\affiliation{12}{School of Civil and Environmental Engineering, Cornell University, Ithaca, NY, USA}
\affiliation{13}{College of Civil Engineering, Tongji University, Shanghai, China}
\affiliation{14}{Independent Researcher, Maryland, 20878, USA}
\affiliation{15}{Zhejiang Institute of Hydraulics and Estuary (Zhejiang Institute of Marine Planning and Design), Hangzhou 310020, China}
\affiliation{16}{Hydro-Climate Extremes Lab (H-CEL), Ghent University, Ghent, Belgium}
\affiliation{17}{Department of Earth, Ocean \& Atmospheric Sciences, University of British Columbia, Vancouver, V6T 1Z4, Canada}
\affiliation{18}{College of Water Sciences, Beijing Normal University, Beijing, China}
\affiliation{19}{Karlsruhe Institute of Technology (KIT), Institute of Water and Environment, Karlsruhe, Germany}
%(repeat as many times as is necessary)

% Corresponding author (used in published mode margin).
% Do not prepend with "SI Corresponding author: " in published version.
\authoraddr{Baoying Shan, Department of Civil and Environmental Engineering (D.I.C.A.), Politecnico di Milano, 20133 Milano MI, Italy, baoying.shan@polimi.it}

% Also define~\correspondingauthor for draft-mode display at page bottom.
% Example:~\correspondingauthor{First Last}{email@address.edu}
\correspondingauthor{Baoying Shan}{baoyingshan0@163.com}

\authorrunninghead{Ignored.} % Ignored in published mode.
\titlerunninghead{Ignored.}  % Ignored in published mode.

%%%%%%%%%%%%%%%%%%%%%%%%%%%%%%%%%%%%%%%%%%%%%%%
% KEY POINTS
%%%%%%%%%%%%%%%%%%%%%%%%%%%%%%%%%%%%%%%%%%%%%%%
%  List up to three key points (at least one is required)
%  Each must be 140 characters or fewer with no special characters or punctuation
%  and must be complete sentences

\keypoints
    {CHA-Flood transforms tacit forecaster expertise into an auditable skill-orchestrated workflow driven by large language models}
    {Skill-guided prior judgment constrains flood peak and volume estimates while improving hydrological scheme selection}
    {All five large language models execute the workflow with comparable accuracy and each outperforms the same workflow without LLM reasoning}

\maketitle

%%%%%%%%%%%%%%%%%%%%%%%%%%%%%%%%%%%%%%%%%%%%%%%
%  ABSTRACT and PLAIN LANGUAGE SUMMARY
%%%%%%%%%%%%%%%%%%%%%%%%%%%%%%%%%%%%%%%%%%%%%%%

\begin{abstract}
Operational flood forecasting depends on tacit forecaster expertise that is difficult to formalize, audit, and transfer.
Although artificial intelligence methods have advanced flood prediction and model-error correction, few studies explicitly represent the tacit expert rules, review checkpoints, and workflow constraints that connect model outputs to operational warning decisions.
Here, we propose Community HydroAgent for Flood Forecasting (CHA-Flood), a skill-orchestrated agent framework that embeds Large Language Models (LLMs) into a model-driven flood forecasting workflow, where each skill encodes explicit rules to bound LLM reasoning. We validated its effectiveness using five state-of-the-art LLMs in the South Yamhill River basin (USA). 
Prior judgment captures observed peak flow and flood volume within 5\% tolerance in 10 and 11 out of 14 events, and 5-fold cross-validation over 129 events yields Pearson correlations of 0.62 and 0.84. 
Building on a high-baseline scheme library (average KGE 0.890), the guided scheme selection further improves KGE by 0.023--0.154.
All five tested LLMs successfully execute the CHA-Flood workflow with comparable judgment accuracy (each spanning 40\%--80\% across cross-validation folds) but with substantial cost differences.
Removing only the language-model narrowing lowers median KGE from 0.723--0.753 to 0.662 for all five LLMs, indicating that the gain comes from LLM reasoning rather than from the choice of LLM.
CHA-Flood does not aim to replace human forecasters; instead, it translates their tacit expertise into an auditable and reproducible workflow. This skill-orchestrated paradigm demonstrates how explicit rule boundaries can guide LLM reasoning to complement physically based simulation in next-generation flood forecasting.

     \begin{plainlanguagesummary}
            Flood forecasts are not produced by hydrological models alone; they also depend on experienced forecasters who judge whether model results make sense for a particular storm. This experience is valuable, but it is often difficult to formalize and transfer. We developed CHA-Flood, a workflow that uses large language models within explicit hydrological rules to make the judgment process more transparent and repeatable. Before forecasting, CHA-Flood groups past floods by magnitude and prepares model parameter schemes for different flood types. When a new flood begins, it estimates plausible ranges for the flood peak and total volume by comparing current rainfall, wetness, and river conditions with similar historical events. Then, it uses those ranges to guide the selection of pre-calibrated hydrological model parameters. The hydrological model then produces the forecast, which is updated as new observations arrive and reviewed by human forecasters before warning information is released. In tests for the South Yamhill River basin, CHA-Flood placed observed flood peaks and volumes within its estimated ranges and improved simulations for several events. This study shows that CHA-Flood can help organize and audit forecaster experience while keeping physical models and human judgment central to flood-warning decisions.
     \end{plainlanguagesummary}
%
% Do not put more abstract text body after plainlanguagesummary!
\end{abstract}

%%%%%%%%%%%%%%%%%%%%%%%%%%%%%%%%%%%%%%%%%%%%%%%
%  BODY TEXT
%%%%%%%%%%%%%%%%%%%%%%%%%%%%%%%%%%%%%%%%%%%%%%%

\section{Introduction}
% [Potentially relevant references have been added in the refs.bib file, please review and use as needed.]

% Subsection 1.1
% author: Gaoyu Liu & Qingyi Yang

% The Importance of Forecaster in Flood Forecasting
Climate change is intensifying the frequency and severity of extreme floods~\cite{allen2002constraints, guan2024human, nearing2024global}. Accurate and timely operational flood forecasting systems are increasingly important for loss mitigation and basin-scale risk management~\cite{fraehr2023supercharging, najafi2024high}. Current forecasting systems are mainly built on hydrological models, e.g., process-based models. By simulating rainfall--runoff processes, they produce preliminary flood hydrographs and support early warning decisions~\cite{jia2026streamflow, mceachran2024parsimonious}.
Guided by hydrological model forecasts, experienced forecasters make informed judgments and issue reliable warnings in operational practice~\cite{pagano2016automation}. However, raw process-based model outputs are rarely adopted directly as the final forecast, due to the spatial heterogeneity of precipitation, inherent uncertainty in meteorological forecast data, and model parameterization errors~\cite{mceachran2025knowledge, tran2026value}. Thus, under the prevailing ``forecasters-in-the-loop'' paradigm, forecasters need to critically evaluate and adjust model-generated forecasts by comprehensively considering meteorological, geophysical, and operational information~\cite{demeritt2013european, emerton2016continental, pagano2016automation}.
For streamflow prediction and flood forecasting, recent comparative evaluations further suggest that forecaster-involved systems consistently outperform standalone machine learning models, including Long Short-Term Memory (LSTM) networks and other deep learning architectures that have shown strong predictive accuracy in rainfall–runoff modeling~\cite{kratzert2018rainfall, frame2022, nearing2024global}, even when precipitation inputs contain substantial errors~\cite{dong2025deep, tran2026value}.
Therefore, the professional judgment of experienced forecasters serves as a second safeguard for the quality of operational flood forecasts, complementing model simulations.

% What are the problems with this paradigm? Difficulty in transferring experience + long training time.

However, this ``forecasters-in-the-loop'' paradigm also exposes a fundamental bottleneck. The adjusting model parameters and reinterpreting simulated hydrographs process by forecasters are inherently tacit and subjective, making such adjustments and their impact on forecast performance difficult to explicitly represent or rigorously evaluate from a scientific perspective~\cite{boyle2000toward, gauch2023defense, stuart2022evolving}.
Furthermore, training a qualified operational forecaster typically requires years of field experience, yet such expertise remains highly personal and difficult to transfer systematically across individuals or institutions~\cite{ladue2018facilitating, tran2026value}.
As a result, the most critical to operational flood forecasting—expert judgment—is the component least amenable to explicit representation, independent validation, and systematic transfer.
This is a major bottleneck for moving flood forecasting toward greater automation and auditability.
Meanwhile, this paradigm faces mounting pressure from rapidly expanding observational networks, real-time decision-making under incomplete information, and flood events that may exceed the historical range under climate change~\cite{kim2024higher, merz2021causes}.

% Subsection 1.2:
% author: Xiaoyi Dong & Qingyi Yang

% How to deal with? AI & DL
Artificial intelligence (AI), particularly deep learning (DL), has already been widely applied in hydrology, including flood forecasting, streamflow simulation, and model error correction~\cite{frame2022,kratzert2018rainfall, nearing2024global, roy2023novel}.
AI systems can process massive datasets at scale without fatigue and reduce subjective bias~\cite{sun2021explore}. Especially, recent advances in physics-informed deep learning have further improved predictive accuracy, generalization, and physical consistency across diverse hydrological conditions~\cite{bindas2024improving, shen2023differentiable, yan2023pcssr}.
However, these deep learning for rainfall–runoff simulation, including physics-informed deep learning model still offer limited interpretability regarding underlying hydrological processes. As a result, these models often function as input–output mappings, which limits trust within the hydrological community~\cite{de2023towards, feng2022differentiable}.
Moreover, AI and DL models can produce predictions inconsistent with established hydrological knowledge, making them difficult to trust in out-of-sample basins~\cite{nearing2021role, read2019process}. Even when such models yield correct outputs, it may remain unclear whether those outputs were obtained for physically valid reasons~\cite{kirchner2006getting, rudin2019stop}.
Accordingly, even when DL models achieve stronger prediction performance than process-based models in practice, their outputs still require expert judgment and verification before operational use.

% LLM and agent-based systems provide new possibilities -- skill
More recently, large language models (LLMs) and agent-based systems have presented new capabilities that extend beyond conventional numerical modeling~\cite{boiko2023autonomous, ma2026autonomous, yao2022react}.
Traditional DL approaches fit input--output relationships, while LLMs enable natural language reasoning, task planning, and tool invocation~\cite{yan2025aquah, pursnani2024hydrosuite, ramirez2024ai}.
Unlike conventional black-box models that produce only numerical outputs, LLMs can articulate the rationale behind each decision in natural language, yielding a human-readable reasoning trace that, when embedded in a structured workflow, becomes inspectable by domain experts.
These capabilities offer the potential to bridge the gap between computational frameworks and operational flood forecasting~\cite{he2025iwms}, not by replacing numerical models, but by supporting scenario-aware, forecaster-like judgment.
However, most LLM applications in hydrology have focused on (1) prompt-based assistance, (2) post-event reporting, (3) question-answering tasks~\cite{pursnani2024hydrosuite}, and (4) single-task applications, such as hydrological model calibration~\cite{li2026hydroagent, zhu2026large}, hydropower scheduling~\cite{luo2025improving}, and hazard narrative extraction~\cite{zhou2025identification}.
Although single-step tool invocation can support specific computations, current LLM-based applications still lack integrated workflow orchestration capabilities for representing and executing the structured, multi-stage reasoning process of operational flood forecasting. % author: Bing & Qingyi Yang
Recent advances in agent systems have introduced skills as an organizational paradigm to address this limitation~\cite{li2026skillsbench, ling2026agent}. A skill encapsulates a reusable, modular capability, typically comprising natural-language instructions, invocation logic, tool-usage procedures, and executable code, thereby enabling an agent to solve recurring classes of problems~\cite{su2026skill}.
This structure enables LLM agents to reason and act within explicit rule boundaries, rather than relying on unconstrained generation.
Flood forecasting is especially well matched to this organization because it is inherently a staged workflow, in which each step has defined inputs and outputs, domain rules, physical constraints, and review boundaries~\cite{nevo2022flood}. 
Existing hydrological agents span a graded autonomy scale, from LLM-based model calibration~\cite{li2026hydroagent, zhu2026large} to systems that configure and run models, analyze outputs, and prepare reports under human oversight~\cite{yan2025aquah, yan2026ai}.
Alongside this expansion of task autonomy, experience-grounded approaches address how knowledge from past events can inform decisions in new warning situations.
In extreme-weather early warning, SIREN retrieves historical cases and automatically distills reusable guidance to support later warning tasks~\cite{ni2026siren}.
What is needed, therefore, is not another black-box predictor, nor further automation of the modeling workflow itself. It is an agent structure that uses LLM reasoning to organize scenario interpretation, deterministic computation, and human review within one executable workflow, and crucially, to represent the tacit expert judgment that precedes the simulation.

% Subsection 1.3:
% author: Bing & Qingyi Yang

% Contribution of this paper
In this study, we present Community HydroAgent for Flood Forecasting (CHA-Flood), a skill-orchestrated framework for integrating LLMs into a hydrological-model-driven flood forecasting workflow.
Rather than replacing the real-time flood forecasting, CHA-Flood is asked to coordinate the workflow that surrounds the hydrological model: scheme preparation, scenario understanding and prior judgment, flood type selection, rolling forecasting, and bulletin publication.
The resulting system is intended to be accurate, auditable, and reproducible, while still allowing language-model reasoning where human expertise is normally required.
The main contributions of this paper are threefold.
First, we propose CHA-Flood, a skill-orchestrated framework that embeds LLMs into a hydrological-model-driven flood forecasting workflow, combining fixed skill boundaries, explicit forecaster review gates, and hydrological computation performed outside the language model.
This represents a new mode of human--AI collaboration in which expert knowledge, deterministic computation, and language-model reasoning are integrated within an auditable operational pipeline.
Second, we formalize forecaster-style analog event judgment as an executable and auditable skill (Step~1), thereby enabling expert reasoning to be represented as a repeatable operational procedure.
Third, we evaluate CHA-Flood across five state-of-the-art LLMs in the primary basin and conduct Step~0--Step~2 experiments in two additional basins (reported in the supplementary material), distinguishing robust workflow execution from basin-dependent predictive skill.

The rest of this paper is organized as follows. Section~2 describes the CHA-Flood framework, the skill-execution boundary, the Step~1 analog event judgment mechanism, and the study area and evaluation protocol. Section~3 presents a workflow case report and the benchmark results. Section~4 discusses their implications, limitations, and way forward, and Section~5 concludes the paper.

\section{Methodology}

\subsection{Overall CHA-Flood Workflow}

CHA-Flood is a skill-orchestrated agent framework for flood forecasting that embeds a language model into a physically grounded workflow rather than replacing that workflow with end-to-end black-box prediction. The system is organized around three distinct role boundaries. The skill layer defines what should be done: it specifies the operational rules, input-output contracts, and quality constraints that govern each forecasting step. The LLM layer defines how to reason: it interprets the event context, coordinates the workflow, and generates explainable text, but it does not directly compute hydrological quantities. The tool layer defines how to compute: it performs deterministic scoring, model calibration, simulation, and artifact generation. This separation is deliberate. In flood forecasting, the language model should not validate its own numerical output, and the executor should not decide what the business logic means. CHA-Flood therefore uses skills to constrain what the language model may request, while ensuring that hydrograph simulations are executed by deterministic hydrological models in the tool layer rather than by the language model itself.

\begin{figure}[htbp]
  \centering
  \includegraphics[trim=0cm 0cm 0cm 0cm,clip,width=\textwidth]{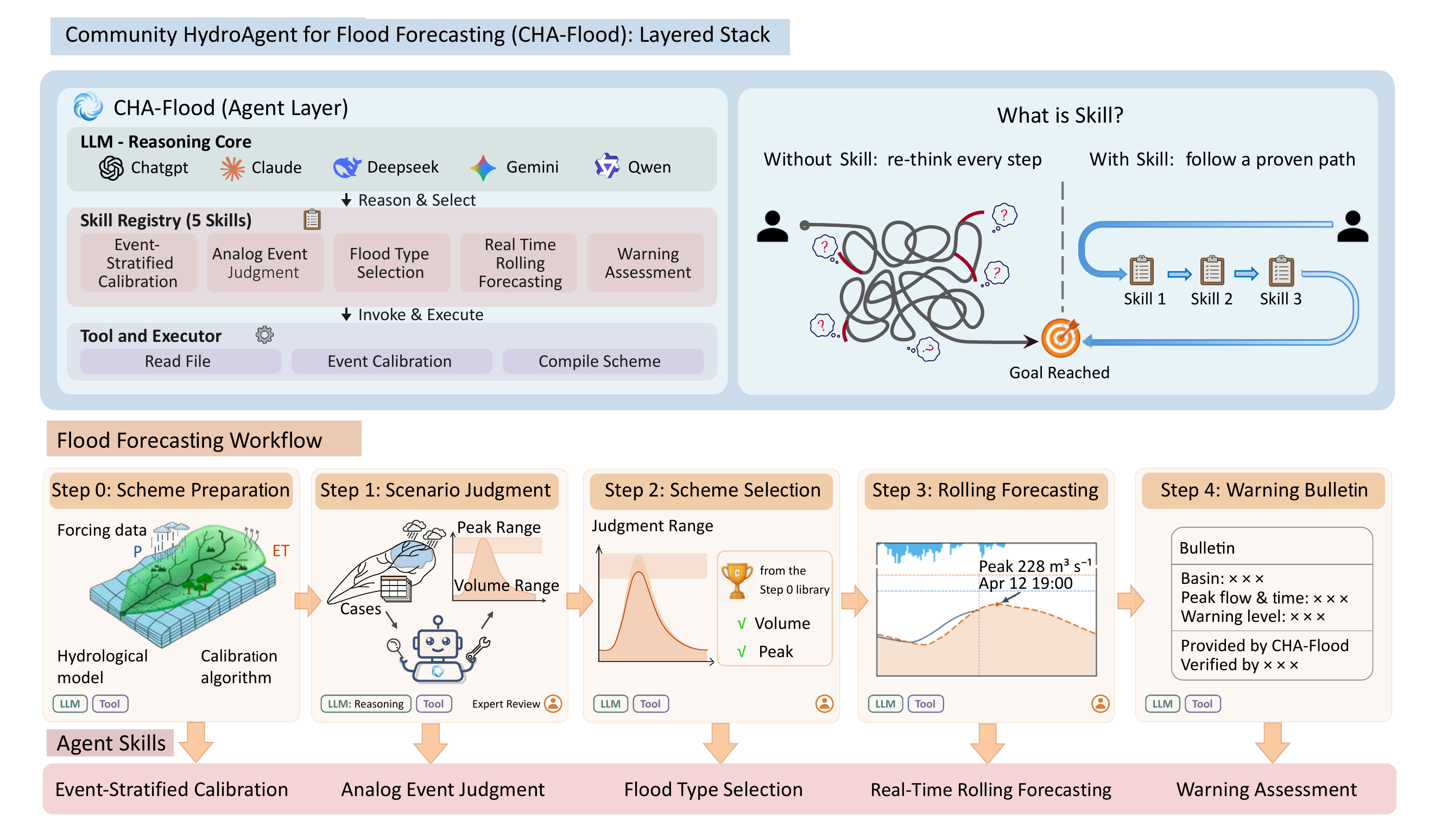}
  \caption{Schematic overview of the CHA-Flood framework.}
  \label{fig:HydroAgent_Framework}
\end{figure}

CHA-Flood workflow includes five steps: (1) scheme preparation, (2) scenario judgment, (3) scheme selection, (4) rolling forecasting, and (5) warning bulletin. Across all steps, structured JSON interfaces (e.g., scheme.json, judgment.json, and forecast.json) transfer information between successive steps and keep the workflow traceable and reproducible (Figure~\ref{fig:HydroAgent_Framework}).
In Step~0, scheme preparation, historical flood events are first classified into distinct groups according to peak flow magnitude (e.g., Type~I--V), with basin-specific thresholds detailed in Section~2.4. Events within each group are then used to independently calibrate the parameters of a hydrological model, the Xinanjiang (XAJ) model~\cite{ren1992xinanjiang}, using the Dynamically Dimensioned Search (DDS) algorithm~\cite{Tolson2007DDS}. Detailed descriptions of the XAJ model and the DDS algorithm are provided in Text~S1 and Text~S2, respectively. The objective function, parameter bounds, and stopping criterion of this calibration are given in Text~S3 and Table~S1.
Calibration yields one XAJ parameter set per flood type. These type-specific parameter sets together form the basin-specific forecasting scheme---a type-indexed parameter library, which provides the basis for the subsequent real-time steps.

\begin{figure}[htbp]
    \centering
    \includegraphics[width=\textwidth]{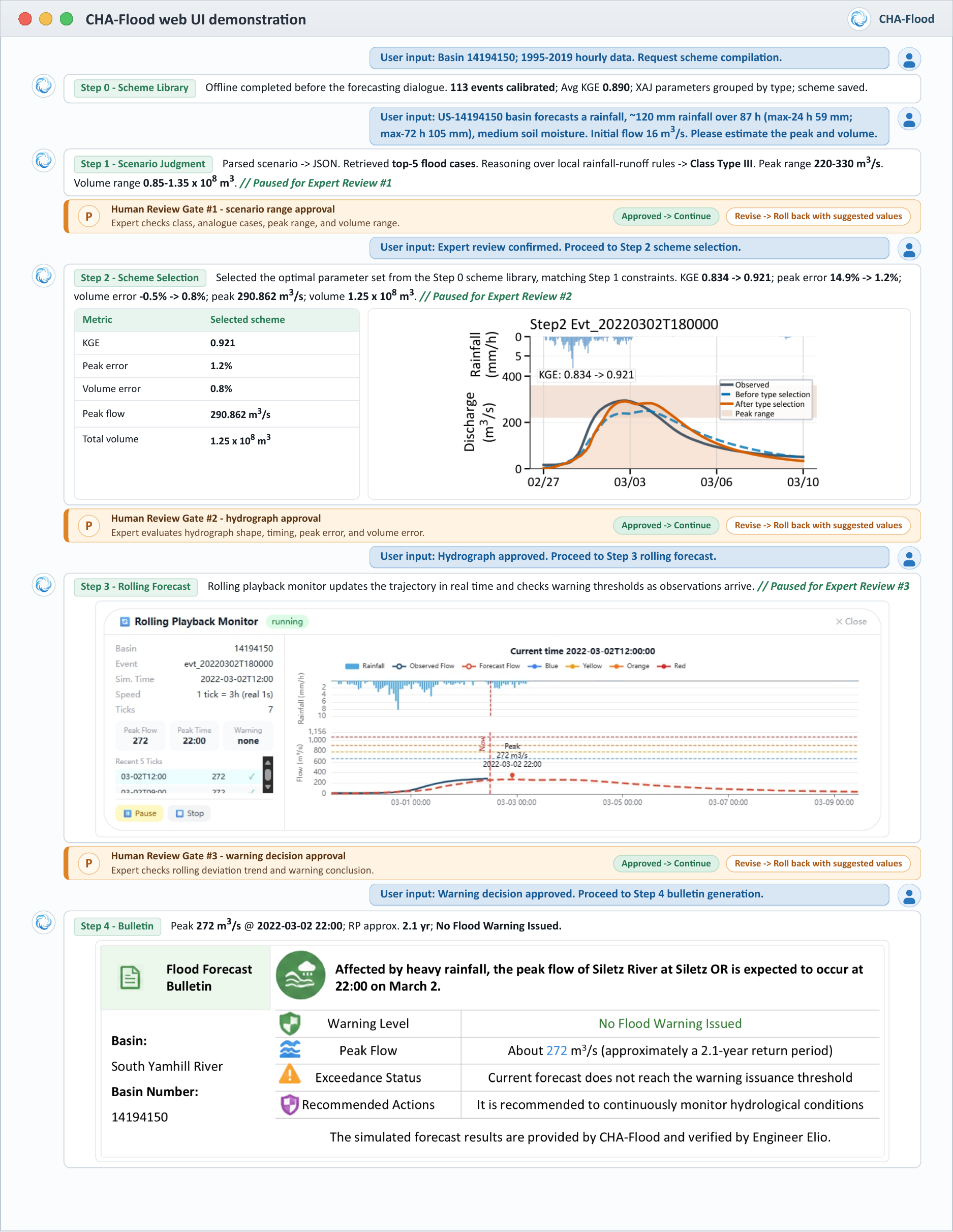}
    \caption{End-to-end workflow of CHA-Flood for the March 2022 flood event on South Yamhill River (basin 14194150), illustrating the Step~0--Step~4 chain and three forecaster-in-the-loop expert review checkpoints.}
    \label{fig:workflow}
\end{figure}

When a potential flood event is forecast, CHA-Flood enters the real-time forecasting pipeline, which comprises four successive steps. Step 1, scenario judgment, is the core validation target of this work. Here, the agent parses the scenario, retrieves similar historical cases, and derives prior intervals for peak flow and flood volume. 
Step~2, scheme selection, retrieves from the Step~0 library the parameter set matching the Step~1 prior ranges and runs the XAJ simulation. The detailed selection rule is summarized in Text~S4.
Step 3, rolling forecasting, updates the simulation as new real-time rainfall and discharge information arrives. Step 4, warning bulletin, generates the formal forecasting report and warning output under human review.
Among these steps, Step~1 constitutes the reasoning core of CHA-Flood, translating forecaster expertise into LLM-driven reasoning through the analog event judgment skill. Steps~0 and~2 
use the event-stratified calibration and flood type selection skills, respectively, and together form the hydrological core in which the hydrological model and the calibration algorithm execute deterministic computation. Steps~3 and~4 form the early-warning and post-processing module, using the real-time rolling forecasting and warning assessment skills, respectively. Each step operates within the scope defined by its corresponding skill. CHA-Flood pauses after Steps~1, 2, and~3 (Expert Reviews~\#1--\#3 in Figure~\ref{fig:workflow}) to await expert approval before advancing, keeping the workflow auditable and reliable.

\subsection{Skill Design and Execution Boundary}\label{sec:skill_design}

In CHA-Flood, a skill is the unit that turns a hydrological task into an executable module. 
A skill is defined as a statically authored, self-contained procedural specification that encodes a complete workflow, including instructions, permitted tools, accessible resources, and a required output structure. It functions as a controlled execution template that constrains and guides the agent's behavior for a specific class of tasks. Each skill is stored in a SKILL.md file.
Following the principle of progressive disclosure, only the skill's lightweight metadata is preloaded into the active prompt. The full body is dynamically loaded on demand when the skill is matched to the current task~\cite{wu2026behavioralintegrityverificationai}. During execution, the active skill boundary is fixed for the run, and the available tools are restricted by the skill's metadata together with runtime tool filtering. Each skill contains four components. First, \textit{Instructions} define the workflow logic and the do-and-don't constraints that the agent must follow. Second, \textit{Tools} define which deterministic actions the agent may invoke, together with the expected input and output schemas. Third, \textit{Resources} define what data sources the skill may access, such as the historical case library, basin observations, or scheme files. Fourth, an \textit{Output Contract} defines the canonical artifact that must be produced when the skill completes. This design keeps the skill boundary explicit and stable across runs, while allowing the underlying language model to vary within the same workflow logic. CHA-Flood thereby separates procedural control from event-specific reasoning and deterministic computation.

Step 1 is the clearest example of this organization. The Step 1 skill requires the agent to parse the event, retrieve similar cases prior to discharge simulation based on antecedent conditions such as rainfall and wetness, compare the current event with the historical record, and apply physical constraints before producing a prior judgment, followed by an explicit human review gate. Its tool set includes \texttt{load\_cases} for case-library overview, \texttt{score\_cases} for similarity computation, \texttt{get\_case\_detail} for retrieving full case records, \texttt{get\_scoring\_rules} for accessing hydrological constraints, and \texttt{save\_prior\_judgment} for contract persistence. Its resources include the case catalog and the Step 0 scheme. Its output contract includes \texttt{judgment.json} for machine consumption and \texttt{report.md} for human review. This structure ensures that the reasoning trace is recorded, the numerical score is reproducible, and the final prior judgment can be passed directly to Step 2.

This separation clarifies the respective responsibilities of each layer and supports auditability, numerical reproducibility, and procedural maintainability. The skill layer defines the procedure, the language model interprets event-specific differences within that procedure, and the tool layer performs deterministic scoring and artifact generation.
As a result, CHA-Flood preserves reproducibility at the execution level while keeping the workflow logic maintainable across runs.
The language model is the only stochastic component of this arrangement. For every result reported below, all LLM calls requested a temperature of 0.1, 
while the setting controlling the range of candidate tokens considered during generation was left at each provider's default and no fixed random seed was specified.
The temperature was not set to zero, doing so generally causes the decoding algorithm to behave greedily by selecting the most likely token at each generation step~\cite{li2025exploring}.
Therefore, 0.1 was adopted as a low-randomness default and was not tuned for this study. The similarity score is deterministic and the prior interval is produced by the language model, so run-to-run variation enters only at Step~1 and is quantified in Section~\ref{sec:benchmark}.

\subsection{Step~1 Analog Event Judgment Skill}

The analog event judgment skill in Step 1 formalizes the judgment process that a human forecaster would normally perform when turning event information into a prior forecast interval. In practice, forecasters combine quantitative event attributes, such as total rainfall, with insights from similar historical cases to estimate plausible ranges of peak flow and flood volume. To make this process explicit, the skill is organized into four computational stages, followed by a final human review gate: scenario parsing, analog retrieval, expert-rule reasoning, and physical constraint validation. The workflow begins with scenario semantic parsing. 
Given a free-text event description, the language model extracts structured variables such as total rainfall, rainfall duration, maximum short-duration rainfall, antecedent wetness class, event type, and initial discharge.
The antecedent wetness class is derived from antecedent precipitation $P_{7d}$, the cumulative areal rainfall over the seven days preceding the event start~\cite{farrick2014soil}. Events are assigned to four classes: dry ($P_{7d} < 20$~mm), medium ($20 \le P_{7d} < 50$~mm), wet ($50 \le P_{7d} < 100$~mm), and saturated ($P_{7d} \ge 100$~mm).
If fields are missing, the model must request clarification rather than generate non-observed values. This requirement is important because the prior judgment is only meaningful when it is anchored in an explicit event description.

First, once the event has been parsed, a six-dimensional similarity retrieval is performed in which the executor computes a weighted score over the historical case library through \texttt{score\_cases} and returns the top five analogs. 
If the fifth- and sixth-ranked cases have the same total similarity score, all top six cases are retained as reference events.
Retrieval of hydrologically analogous events has also been used to improve forecasts of large floods in data-driven settings~\cite{lin2023exploring}. 
The scoring dimensions represent the main controls on flood response: total rainfall, rainfall duration, antecedent wetness, event type, maximum short-duration rainfall, and initial discharge. The 13-point score assigns up to 3 points to total rainfall, reflecting its dominant influence on runoff magnitude, and up to 2 points to each of the remaining five dimensions. The scoring weights, set by the operational forecasters among the authors, formalize basin-specific forecaster judgment. We evaluated weight perturbations, information ablation, and downstream Step~2 responses across the 14 validation events, with KGE differences calculated relative to the baseline results (Text~S5 and Table~S2). Most weight-perturbation runs retained the parameter scheme selected under the original weighting, whereas some information-ablation conditions altered analog retrieval and prior judgment, leading to downstream scheme changes and reduced KGE in individual events. Continuous variables are scored using relative-difference functions, whereas antecedent wetness and event type are scored using ordinal and categorical similarity rules, respectively.
The use of a continuous rather than categorical score makes the retrieval more discriminative and helps improve the ranking among closely related hydrological events. This retrieval step instantiates the case-based example introduced in Section~\ref{sec:skill_design}. Supplementary Text S6 and Table S3 summarize the scoring rule used in Step 1.
The number of retrieved analogs $N$ is a configurable skill parameter and was set to five in this study. Figure~S1 tests this choice: among $N$ = 3, 5, 7 and 10, five gave the highest median Kling--Gupta efficiency (KGE) and the narrowest interquartile range. 
However, the difference between $N = 5$ and $N = 7$ is smaller than the spread of eleven repeats of that configuration itself and cannot be separated from run-to-run variation. Within the range tested, $N$ can therefore be adjusted to the needs of a given application, with five recommended as the default.

Second, the retrieved analogs then enter an expert-rule reasoning stage. The language model loads the full case records with \texttt{get\_case\_detail} and accesses hydrological constraints with \texttt{get\_scoring\_rules} before applying expert judgment to correct the raw similarity result. The reasoning is organized into two complementary parts: analog correction and boundary handling. Analog correction accounts for differences in total rainfall, antecedent wetness, rainfall intensity, storm duration, and basin-scale runoff controls. For example, short-duration intense rainfall is treated as more likely to produce sharper peaks, whereas persistent low-intensity rainfall is treated as more likely to produce attenuated hydrographs \cite{breinl2021understanding}. Boundary handling then uses an anchor case to constrain the upper bound when a close analog exists and falls back to controlled extrapolation when the target event lies outside the support of the case library. These rules allow the skill to refine the similarity score into a physically plausible prior interval rather than relying solely on nearest-neighbor matching. The detailed rule definitions are summarized in Supplementary Text S7 and Table S4.

Third, before the prior judgment is finalized, a physical red-line validation verifies that the proposed interval satisfies a runoff-coefficient constraint $R \in [0.05, 1.1]$.
The upper bound allows for uncertainty in rainfall estimates~\cite{tran2025camelsh} and additional water inputs beyond rainfall, such as episodic snowmelt~\cite{csen2006comparative}. The interval must also exclude obvious hydrological inconsistencies, such as a forecast peak below the initial discharge.
If a proposed interval fails a physical check, the skill tightens or adjusts the interval and records the reason in the report. The validated prior-judgment package includes the peak flow and flood volume intervals, similar cases, flood type, confidence, and a reasoning summary. 
Fourth, this package is written to \texttt{judgment.json} for downstream consumption and to \texttt{report.md} for human review. The final forecaster-in-the-loop confirmation then acts as an explicit gate before the judgment is persisted and passed to Step 2. This stage is not a separate computational step, but rather the point at which the prior judgment becomes eligible for downstream scheme selection. The same design pattern extends naturally to the other skills, whose full instructions and tool interfaces follow the same skill-based structure.

To further assess the generalization of Step 1, we use a stratified 5-fold cross-validation over 129 historical flood events spanning 1995--2024. Events are grouped by flood class (Type~I--V) and distributed across folds in a round-robin manner so that each fold preserves the overall class distribution; within each fold, the agent is evaluated on held-out events and is only allowed to retrieve cases from the corresponding reference pool. This design tests whether the Step 1 analog event judgment skill can recover physically reasonable intervals when the target event is excluded from the retrieval pool. This validation is reported separately from the 14-event benchmark used to demonstrate workflow execution.

\subsection{Study Area and Data}
% author:Jiale Guo
% review and updated by Baoying Shan 2026.05.11

CHA-Flood is tested in the South Yamhill River basin, with USGS gauge 14194150 at McMinnville, Oregon, United States, used as the outlet. This basin serves as a demonstration testbed rather than a fixed domain of application: the same workflow can be transferred to other basins once local observations, model schemes, warning thresholds, and forecaster rules are converted into basin-specific skills and case libraries. The South Yamhill basin was selected because its humid runoff response is consistent with the main assumptions of the XAJ model, especially saturation-excess runoff generation, with a high runoff coefficient, a predominantly rainfall-driven regime, and no major reservoir regulation.
The analysis used hourly hydrometeorological records from the CAMELSH data set for 1995--2024, including streamflow observations for this gauge~\cite{tran2025camelshdata, tran2025camelsh} (Zenodo DOI: \url{https://doi.org/10.5281/zenodo.16763144}; \textit{Scientific Data} article DOI: \url{https://doi.org/10.1038/s41597-025-05612-6}).

Flood events were extracted from the hourly record using a peak-over-threshold procedure. For each event, we calculated total rainfall, rainfall duration, maximum short-duration rainfall, antecedent wetness, initial discharge, peak discharge, and flood volume. Events were assigned to five basin-specific flood types according to observed peak discharge: Type~I ($Q_p \geq 490$~m$^3$~s$^{-1}$; 14 events), Type~II ($350 \leq Q_p < 490$~m$^3$~s$^{-1}$; 23 events), Type~III ($260 \leq Q_p < 350$~m$^3$~s$^{-1}$; 22 events), Type~IV ($205 \leq Q_p < 260$~m$^3$~s$^{-1}$; 26 events), and Type~V ($175 \leq Q_p < 205$~m$^3$~s$^{-1}$; 28 events). 
The flood types and their thresholds are basin specific and can be adjusted in the prompt. Here they were set to balance the number of events across types.

The event archive contains 129 historical flood events distributed across these five types. A subset of 113 events from 1995--2019 was used as the historical case library and as the calibration set for the Step~0 XAJ parameter schemes. The 14 events from 2020--2024 were held out for independent validation of Step~1 scenario judgment and Step~2 scheme selection. 
Two flood events in the validation period were excluded from the analysis (N.15\_20240129 and N.16\_20241229). Their hydrographs are shown in Figure S2 of the Supporting Information. We excluded the Event N.15\_20240129 since its discharge series does not represent a complete flood event: the hydrograph remains nearly flat throughout the period and corresponds to the recession of the preceding event rather than an independent flood. Event N.16\_20241229 was excluded because our discharge record ends on 31 December 2024, whereas the recession of this event extends into 2025, leaving the hydrograph truncated.
The 129-event archive was also used for the stratified 5-fold cross-validation described above.

South Yamhill River was retained as the primary study basin for the complete workflow demonstration, cross-validation, ablation, and multi-LLM experiments. To provide supplementary evidence of cross-basin portability, the shared Step~0--Step~2 workflow was additionally applied to Smith River near Crescent City, California, United States, and Mission River at Refugio, Texas, United States. These basins represent humid and semi-arid hydroclimatic conditions, respectively. For both basins, we used the same 1995--2019 calibration period and 2020--2024 validation period. Their basin characteristics, flood classifications, calibration results, and event-level Step~1--Step~2 results are reported in Text~S8, Figures~S3--S4, and Tables~S5--S7 of the Supporting Information.

\section{Results}

\subsection{CHA-Flood-Driven Flood Forecasting Results}
% case report
The capability of CHA-Flood to orchestrate the full Step~0--Step~4 chain is illustrated with a March 2022 flood event (N.6\textunderscore20220302). This event is selected for its representativeness of a moderate-magnitude storm (max-24~h 59.0~mm), medium antecedent wetness ($P_{7d} = 22.2$~mm), and low initial discharge of 16~$\mathrm{m^{3}/s}$, whose long duration naturally traverses all five workflow stages in a single trace. Prior to the event, Step~0 had already produced a five-class parameter library (Type~I--V, average KGE 0.890 across 113 calibrated events) for the basin (Figure~2). Since event-type stratification is not standard practice in operational hydrology, a complementary experiment was conducted in which a single unified parameter set was calibrated across all 113 events and benchmarked against the five-type scheme; results are reported in Figure S5.

The forecaster initiated Step~1 by submitting a meteorological scenario to CHA-Flood~(Figure~2): 120~mm of areal rainfall over 87 effective hours (mean intensity 1.4~mm/h, max-24~h 59.0~mm), medium antecedent wetness ($P_{7d} = 22.2$~mm), and an initial channel flow of 16.0~$\mathrm{m^{3}/s}$. Retrieving analogous historical cases, Step~1 returned a Type~III class judgment with a peak range of 220--330~$\mathrm{m^{3}/s}$, a volume range of 0.85--1.35~$\times 10^{8}$~$\mathrm{m^{3}}$, medium confidence, and a normal risk flag. The observed peak of 294.49~$\mathrm{m^{3}/s}$ falls within this range, confirming successful scenario recognition. The workflow paused at Expert Review~\#1 (Figure~2), prompting the duty engineer to either approve the results or request a revision which triggers a rollback with suggested values. Only upon confirming the physical reasonableness did the chain advance to Step~2. 

In Step~2, CHA-Flood selected, rather than calibrated, the parameter set for this event. The deterministic procedure first identified the initial flood type from the midpoint of the Step~1 peak interval (Type~III). It then ran one XAJ simulation for each of the five pre-calibrated parameter sets in the Step~0 library, using forecast rainfall and pre-event model states. Finally, a fixed selection rule was applied: among the sets whose simulated peak and flood volume both fall within the Step~1 prior intervals, the set with simulated volume closest to the midpoint of the prior volume interval was chosen. For this event, the Type~II set satisfied both prior intervals (simulated peak 290.9~$\mathrm{m^{3}/s}$, volume $1.25~\times 10^{8}$~$\mathrm{m^{3}}$) and was selected, so the forecast configuration switched from the initially identified Type~III set (\textit{sm} = 188.48~mm, \textit{kg} = 0.0764, \textit{b} = 1.337, \textit{l} = 12~h, \textit{cs} = 0.923) to the raw Type~II set (\textit{sm} = 110.32~mm, \textit{kg} = 0.3698, \textit{b} = 1.082, \textit{l} = 18~h, \textit{cs} = 0.935). No parameter value was modified during the event, both sets are untouched Step~0 calibration products, and the observed hydrograph played no role in the selection. Evaluated post hoc against observations, which are not available to the selection rule, peak error decreased from $-14.9\%$ (initial Type~III set) to $-1.2\%$ (selected Type~II set), volume error changed from $-0.5\%$ to $+0.8\%$, and KGE rose from 0.834 to 0.921; the per-set KGE values are reported as diagnostics only and do not enter the selection. Expert Review~\#2 validated the hydrograph shape before rolling forecasting was authorized.

In Step~3, CHA-Flood assimilated incoming observed flows and rolled the forecast forward in real time, and at Expert Review~\#3 the forecaster approved the warning decision (Figure~2). Once the event receded, Step~4 compiled the bulletin: peak time 2022-03-02 22:00, return period approximately 2.1 years, warning level ``No Flood Warning Issued'', with recommended continued monitoring of hydrological conditions. The following section evaluates Step~1 generalization across all held-out events.

\subsection{Validation of Prior Judgment and Scheme Selection}\label{sec:PriorJudgment_validation}

\begin{figure}[htbp]
  \centering
  \includegraphics[width=\textwidth]{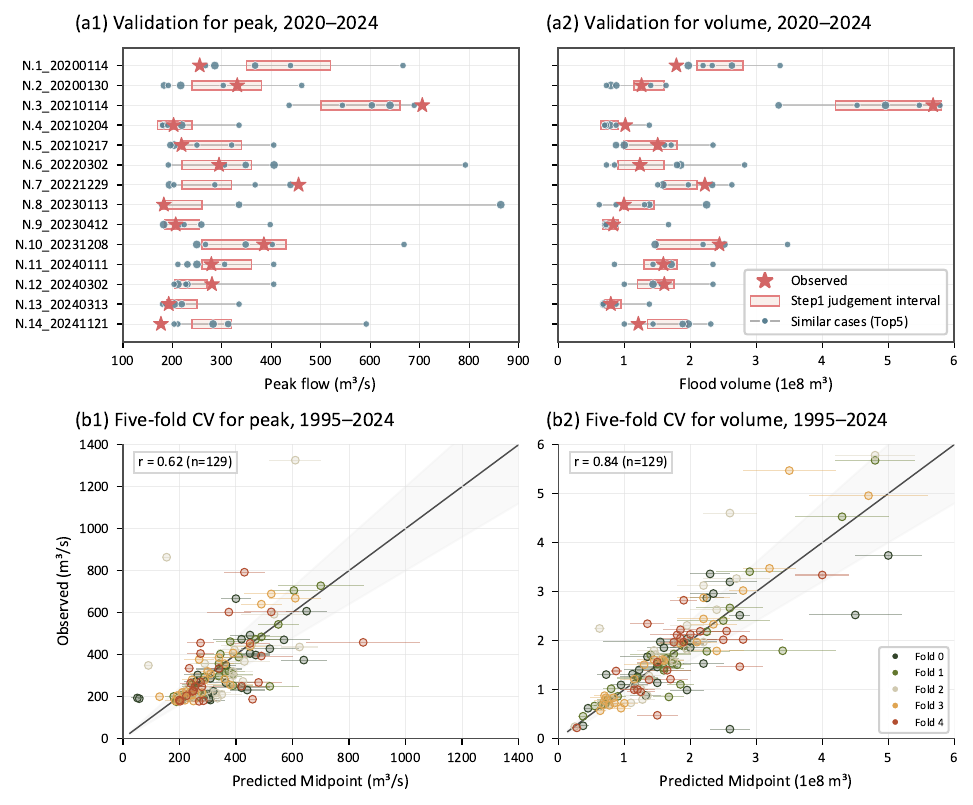}
  \caption{CHA-Flood performance for Step~1 scenario judgment using GPT-5.4.
  (a1) and (a2) show per-event comparison across 14 flood events during the validation period (2020--2024) of observed values (stars), Step~1 judgment ranges (boxes), and Top5 similar historical cases (dots with range whiskers) for peak flow and flood volume, respectively.
  (b1) and (b2) show 5-fold cross-validation results (1995--2024, 129 cases in total) for peak flow and flood volume, respectively. Predicted range midpoints are plotted against observed values, with a 1:1   reference line and $\pm$20\% band. Pearson correlation coefficients are annotated in each panel.}
  \label{fig:fig_step1_combined}
\end{figure}

To validate the performance of Step 1 scenario judgment, we apply CHA-Flood to the selected river basin and evaluate the accuracy of judgment results for peak flow and flood volume across 14 flood events during the validation period (2020--2024) based on historical flood records from 1995 to 2024, as shown in Figure~\ref{fig:fig_step1_combined}a. These results are performed using GPT-5.4 as the LLM backbone, which was one of the most powerful models available at the time of this study. The performance of other LLMs is shown in Figure~S6 and is discussed in Section~\ref{sec:benchmark}. Figure~\ref{fig:fig_step1_combined}a shows the judgment ranges of peak flow and flood volume for each event, as inferred from the five most similar historical cases retrieved by the Step~1 skill. 
The results show that the judgment ranges encompass the observed values in most cases, with hit rates of 10 and 11 out of 14 events for peak flow and flood volume within a 5\% tolerance margin, respectively. 
A judgment is counted as a hit when the observation falls within the predicted interval or lies outside it by no more than 5\% of the observed value. This definition applies to all hit rates reported in this paper.
All judgment ranges are narrower than the spread of the Top-5 similar historical cases, reflecting the refinement introduced by LLM reasoning, with a median interval width of 39\% of the observed value for peak flow and 35\% for flood volume. 
This indicates that the Step 1 skill can effectively constrain the predicted intervals based on the retrieved cases. Two cases (N.1\_20200114 and N.14\_20241121) fall outside the predicted ranges for both peak flow and flood volume, where the LLM overestimated the flood severity. Both events are low-magnitude floods (Type~IV and Type~V, respectively), resulting in relatively mild flood signals that are inherently more difficult to estimate accurately.

To further evaluate the generalization capability of Step 1 scenario judgment, we perform stratified 5-fold cross-validation on 129 historical flood events (1995--2024), as shown in Figure~\ref{fig:fig_step1_combined}b. The results show that the performance across each fold is comparable, with Pearson correlation coefficients ($r$) of 0.62 and 0.84 between observed values and predicted midpoints for peak flow and flood volume, respectively. The judgment performance for flood volume is better than that for peak flow, consistent with the validation results in Figure~\ref{fig:fig_step1_combined}a. One extreme case (evt\_19960209) exhibits a peak flow of 1326 m$^3$/s, nearly twice the maximum peak flow among all other cases. This event is substantially underestimated by the LLM, as its characteristics have no close analogs in the historical record, making accurate judgment inherently challenging.

Following the determination of judgment ranges in Step 1, Step 2 selects the optimal parameter set from the pre-calibrated scheme library (Step 0) to produce the forecast hydrograph. The full results of all 14 validation events are summarized in Table~S8. Among these, 11 out of 14 events yield KGE values after Step 2 scheme selection equal to or greater than those obtained by directly applying the default flood type classification based on the Step~1 prior judgment ranges, with KGE improvements ranging from 0.023 to 0.154 for the five events where the scheme was updated. The simulated peak flow and flood volume after scheme selection fall within the Step 1 judgment ranges (with a 5\% tolerance margin) for 14 and 13 out of 14 events, respectively, confirming that the selected parameter sets produce forecasts consistent with the prior judgment.

\begin{figure}[htbp]
  \centering
  \includegraphics[width=\textwidth]{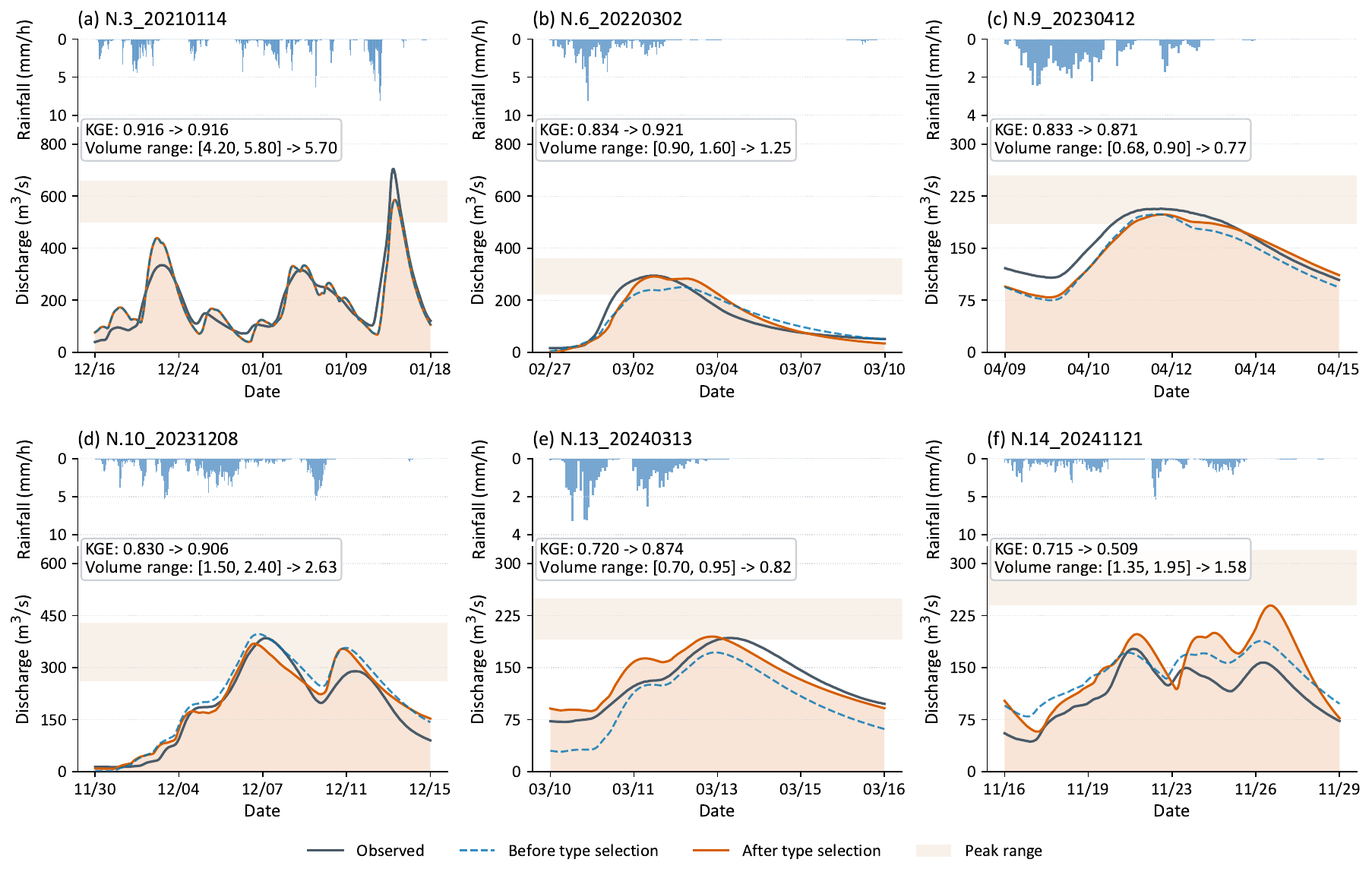}
  \caption{Comparison of simulated flood hydrographs before and after Step 2 scheme selection for six representative validation events: (a) event N.3\_20210114, (b) event N.6\_20220302, (c) event N.9\_20230412, (d) event N.10\_20231208, (e) event N.13\_20240313, and (f) event N.14\_20241121. Blue bars represent hourly rainfall; black, dashed blue, and orange lines represent observed discharge, simulated discharge before scheme selection, and simulated discharge after scheme selection, respectively. The shaded orange band indicates the peak flow range identified in Step 1.}
  \label{fig:SelectedEvents}
\end{figure}

Figure~\ref{fig:SelectedEvents} presents six representative events selected to illustrate a range of outcomes, including cases with substantial KGE improvement (N.6, N.9, N.10, N.13), an unchanged scheme where the default was already optimal (N.3), and a case where Step 1 misjudgment led to degraded performance (N.14). For event N.13\_20240313 (Figure~\ref{fig:SelectedEvents}e), the default simulation (dashed blue) substantially underestimates the peak flow, whereas the scheme selected by Step 2 (orange) closely reproduces both the peak magnitude and the overall hydrograph shape, improving the KGE from 0.720 to 0.874. In contrast, for event N.3\_20210114 (Figure~\ref{fig:SelectedEvents}a), the KGE remains unchanged at 0.916, indicating that Step 2 correctly retains the original scheme when it is already the best available option.

Three events (N.1, N.8, and N.14) exhibit decreased KGE after scheme selection. N.8 and N.14 are Type V floods with observed peak flows only slightly above the minimum flood threshold. N.1 is a low-magnitude Type IV flood, and all three exhibit relatively weak flood signals. For these events, Step 1 tends to overestimate the judgment ranges, as illustrated by the elevated shaded band in Figure~\ref{fig:SelectedEvents}f. This upward bias directs Step 2 to select a parameter set that generates higher peak flows and flood volumes, resulting in overestimated hydrographs that diverge from the observed discharge. This error propagation from Step 1 to Step 2 highlights that the accuracy of Step 1 judgment is a critical prerequisite for the overall forecasting performance of CHA-Flood.

Supplementary cross-basin experiments further examined whether the same Step~0--Step~2 workflow could be executed without basin-specific code duplication. 
For the humid Smith River basin, the Step~1 peak-flow and flood-volume hit rates, including the 5\% near-hit tolerance, were both 57.1\% (12 of 21 events). Median KGE decreased from 0.633 before Step~2 scheme selection to 0.561 afterward, indicating that errors in Step~1 judgment can propagate into Step~2 scheme selection and degrade downstream forecast performance. For the semi-arid Mission River basin, the corresponding hit rates were 0\% (0 of 8 events) and 25.0\% (2 of 8 events), respectively. Median KGE improved from $-1.591$ to $-1.170$, but remained poor in absolute terms. These results support workflow portability but also show that the specific configuration reported in this study, including the variables, weights, and scoring criteria of the six-dimensional similarity-retrieval method, the expert-derived reasoning rules, and the XAJ model, cannot be assumed to generalize across basins without basin-specific adaptation. Predictive skill remains dependent on local hydrological knowledge and model suitability. Complete experimental settings and event-level results are provided in Text~S8, Figures~S3--S4, and Tables~S5--S7.

\subsection{Ablation of the LLM-Derived Prior Judgment}\label{sec:ablation}
\begin{figure}[htbp]
  \centering
  \includegraphics[width=\textwidth]{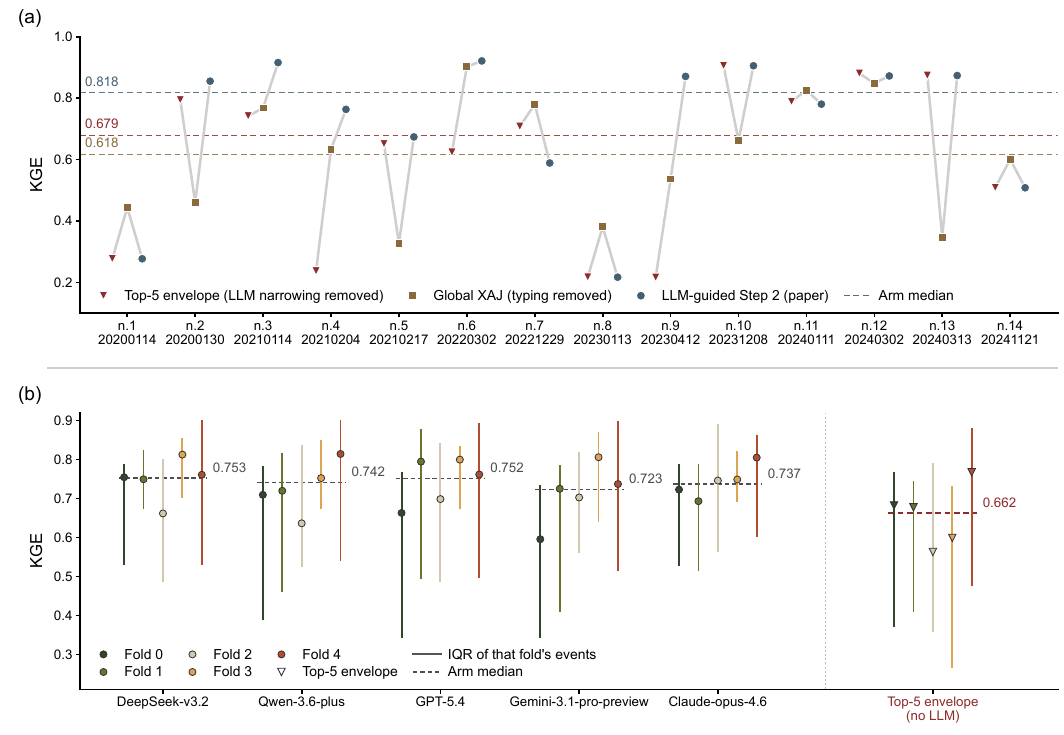}
  \caption{Comparison of CHA-Flood performance with two ablations that remove the language model.
  (a) Per-event Kling--Gupta efficiency (KGE) after Step~2 for the 14 held-out validation floods. Each column is one event, showing CHA-Flood Step~2 (blue circle) against two ablations: the minimum-to-maximum envelope of the Top-5 similar events (red triangle), and a single XAJ parameter set without type stratification, prior interval, or scheme selection (brown square). Dashed lines with the adjacent numeral show each arm's median KGE over the 14 events.
  (b) The same narrowing ablation on the 129-event stratified 5-fold benchmark, for all five LLMs. The points show per-fold median KGE, and the vertical bars show the interquartile range of each fold's 20--28 test events. The dashed lines with the adjacent numeral show each arm's pooled median KGE over all five folds.}
  \label{fig:ablation}
\end{figure}

To isolate what the LLM contributes, we removed its reasoning from the CHA-Flood workflow in two ways and re-ran Step~2 on the same 14 validation events (Figure~\ref{fig:ablation}a). 
The Top-5 envelope arm keeps the retrieval, the five parameter sets calibrated in Step~0, and the Step~2 selection rule. It replaces only the judgment interval with the minimum-to-maximum envelope of the Top-5 similar events. 
That envelope is 1.58 to 8.56 times wider than the judgment interval for peak flow, and 1.53 to 4.31 times wider for flood volume. 
The Global XAJ arm removes type stratification, and with it the prior interval and scheme selection, leaving the standard XAJ model with a single parameter set. Median KGE over the 14 events was 0.818 for CHA-Flood Step~2, against 0.679 for the Top-5 envelope arm and 0.618 for the Global XAJ arm. 
This per-event pattern is consistent with Section~\ref{sec:PriorJudgment_validation}: the accuracy of the prior judgment is critical to the performance of CHA-Flood. 
In the Top-5 envelope arm, the wider interval does not provide better guidance for scheme selection. It wins clearly only on N.7, where the observed peak of 455~$\mathrm{m^{3}/s}$ lay above the judged 220--320~$\mathrm{m^{3}/s}$. There, the KGE of CHA-Flood Step~2 fell to 0.589, below both ablation arms. 
Conversely, for N.9 the judgment interval of 185--255~$\mathrm{m^{3}/s}$ enclosed the observed peak of 207~$\mathrm{m^{3}/s}$ and selected Type~III, whereas the 3.1 times wider envelope admitted Type~I. KGE was 0.871 for CHA-Flood Step~2, against 0.216 for the Top-5 envelope arm and 0.537 for the Global XAJ arm. 
Every loss of CHA-Flood Step~2 larger than 0.05 KGE fell on an event where the Step~1 interval missed the observation or sat at its edge (N.1, N.7, N.8, and N.14).

We repeated the ablation on the 129-event stratified 5-fold benchmark for each of five state-of-the-art LLMs (Figure~\ref{fig:ablation}b). The five models are DeepSeek-v3.2, Qwen-3.6-plus, GPT-5.4, Gemini-3.1-pro-preview, and Claude-opus-4.6, and all were accessed on April 22, 2026. 
Event-stage XAJ parameters were re-calibrated within each fold using only that fold's training events. Pooled median KGE ranged from 0.723 to 0.753 across the five models, against 0.662 for the Top-5 envelope arm. 
No model consistently outperformed the others. The models differed more in the spread of per-event skill than in its center. The pooled interquartile range of KGE was 0.23 for Claude-opus-4.6 and 0.27 for DeepSeek-v3.2, but 0.35 for Gemini-3.1-pro-preview and 0.38 for GPT-5.4. All were narrower than the 0.42 of the envelope arm. 
We did not compare interval hit rates, because the wider envelope would score higher by construction without forecasting better. The informative test is therefore the simulation downstream of the interval. 

\subsection{Cross-LLM Benchmark and Within-LLM Stability}\label{sec:benchmark}
\begin{figure}[!htbp]
  \centering
  \includegraphics[width=0.95\textwidth]{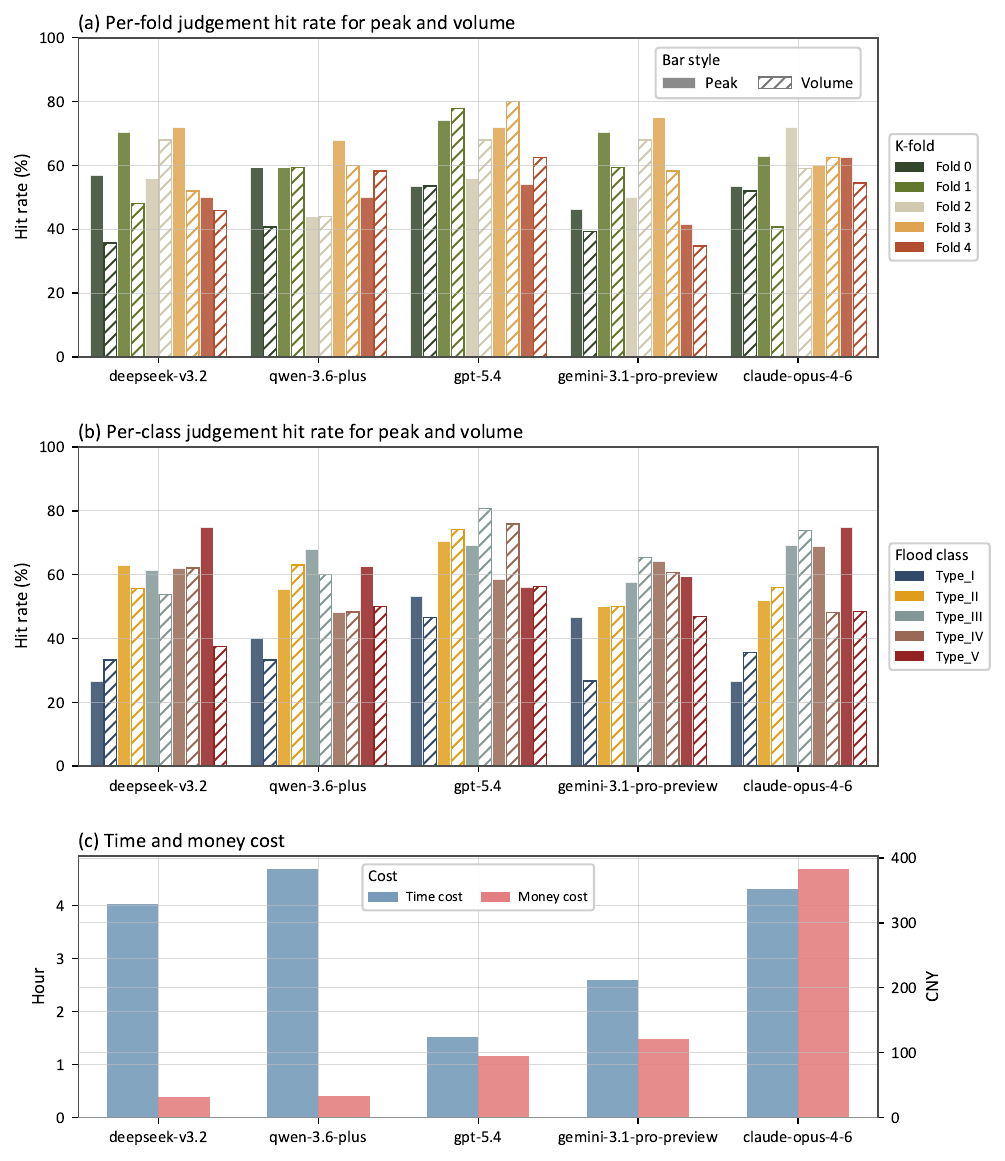}
  \caption{Performance and cost comparison of five LLMs (DeepSeek-v3.2, Qwen-3.6-plus, GPT-5.4, Gemini-3.1-pro-preview, and Claude-opus-4.6) in the Step~1 stratified 5-fold cross-validation over 129 flood events (1995--2024).
  (a) Per-fold hit rates across folds 0--4, with fold sample sizes of 28, 27, 25, 25, and 24 events, respectively.
  (b) Per-class hit rates across flood classes Type~I--V, with corresponding event counts of 15, 27, 26, 29, and 32.
  (c) Running time (hours, left axis) and monetary cost (CNY, right axis) for the full benchmark.
  In (a) and (b), solid bars denote peak-flow hit rate and hatched bars denote flood-volume hit rate; color encodes the LLM. A hit is defined in Section~\ref{sec:PriorJudgment_validation}.}
  \label{fig:LLMs_kfold_performance_costs}
\end{figure}

% Cross-LLM benchmark (NEW)
The portability of the skill-orchestrated design is supported by a stratified 5-fold cross-validation benchmark of Step~1 across the same five LLMs (Figure~\ref{fig:LLMs_kfold_performance_costs}). 
All five LLMs operate within the same Step~1 skill and achieve broadly comparable hit rates (Figure~\ref{fig:LLMs_kfold_performance_costs}a, each spanning 40--80\% across cross-validation folds), with GPT-5.4 yielding the highest flood-volume hit rates (74--80\% across Folds~1--3) but no model winning by a clear margin.
These hit rates are achieved with intervals of similar width across the five LLMs. Median interval widths relative to observed values range from 23\% to 31\% for peak flow and from 20\% to 32\% for flood volume.
This supports the central design claim that the workflow, rather than the choice of LLM, governs forecasting quality. In contrast, Figure~\ref{fig:LLMs_kfold_performance_costs}c reveals substantial cost dispersion that is largely decoupled from accuracy: GPT-5.4 completes the full benchmark in 1.5~h, whereas Claude-opus-4.6 and Qwen-3.6-plus require 4.3 and 4.7~h, respectively, and monetary cost spans more than an order of magnitude, from approximately 30~CNY (DeepSeek-v3.2) to 390~CNY (Claude-opus-4.6, roughly 4$\times$ that of GPT-5.4 and 13$\times$ that of DeepSeek). These results suggest that, within the accuracy range observed in this benchmark, budget and latency become first-order considerations alongside accuracy when selecting an LLM for this workflow. Per-class hit rates (Figure~\ref{fig:LLMs_kfold_performance_costs}b) further show that all five LLMs perform worst on Type~I floods (27--53\%), pointing to a shared limitation that is examined below.
% move from the supplementary to here

% Within-LLM stability
Beyond cross-model comparison, we examined the within-model stability of the Step~1 skill by repeating the prior-judgment task ten times for each of the 14 validation events using GPT-5.4 (Figure~S7). 
Across the ten runs, the interval midpoint moved by a median of 16\% of the observed peak flow and 13\% of the observed flood volume (maxima 53\% and 29\%, N.10\_20231208). This movement is about one sixth of the Top-5 envelope width for both quantities. 
Each of the 140 Step~1 outputs from the GPT-5.4 repeats was then carried forward as the input to the deterministic Step~2 selection. All 140 runs completed successfully and produced a forecast, and the published intervals of Table~S8 reproduced their reported KGE exactly. 
Seven of the 14 events selected the same parameter set in every run and therefore produced identical forecasts. For the other seven events, the difference between the highest and lowest KGE across the ten runs was 0.11 to 0.72. The largest difference, for N.5\_20210217, arose because two runs admitted the Type~I set and KGE fell from 0.67 to $-0.05$, a class-boundary effect of the selection rule rather than a large interval shift.
Importantly, the two events (N.1\_20200114, N.14\_20241121) that fall outside the predicted ranges in Section~\ref{sec:PriorJudgment_validation} also fall outside the ranges across all ten repeats, confirming that their overestimation reflects a systematic limitation at the tail of the case library rather than sampling variance.

% Cross-LLM stability (Reviewer 2, Major 2.9)
Given the rapid turnover of commercial LLM versions, we extended the repeated-run evaluation to examine the generality of our findings (Figure~S8).
For each validation event, we executed Steps~1--4 ten times with each of the other four LLMs: DeepSeek-v3.2, Qwen-3.6-plus, Gemini-3.1-pro-preview, and Claude-opus-4.6.
We also tested eight newer models with three repeats each: GPT-6-Astra, Claude-Fable-5.1, Gemini-3.6-Flash, Qwen3.8-Max, DeepSeek-V4-Pro, Kimi-K3, Grok-4.6, and GLM-5.3.
These additional runs were conducted on September 17--18, 2026, using the same 14 validation events.
All 13 LLMs completed the workflow, supporting its applicability across the models tested.
With the offline Step~0 parameter library and event inputs fixed, variation in Step~2 KGE arose from Step~1 prior judgments propagated through deterministic scheme selection.
The median across events of the per-event median KGE was 0.79 for GPT-5.4 and 0.78 for DeepSeek-v3.2.
The corresponding values were 0.71 for Qwen-3.6-plus and Claude-opus-4.6, and 0.69 for Gemini-3.1-pro-preview.
Central skill was broadly similar across the five models, but repeatability differed.
GPT-5.4 selected the same parameter set in all ten runs for seven events, compared with zero to three events for the other four models.
The median per-event KGE range was 0.06 for GPT-5.4 and 0.17--0.28 for the others, with the largest ranges at class-boundary events N.5\_20210217 and N.13\_20240313.
Gemini-3.1-pro-preview showed the largest median peak-interval midpoint range across repeats, at 47\% of observed peak flow, compared with 16--40\% for the other four models.
For the eight newer models, central KGE ranged from 0.67 to 0.80, although three repeats provide limited evidence of stability.
Qwen3.8-Max and GPT-5.4 had the highest central KGE across all 13 models, but no model consistently outperformed the others across events.
Together, these results support the applicability of CHA-Flood across LLMs, while showing that forecast skill and repeatability remain model dependent.

\section{Discussion and Way Forward}\label{sec:discussion}

\subsection{Discussion}

This paper developed a framework to embed large language models into the flood-forecasting workflow and showed that all five tested LLMs could support forecasts that captured observed peak flow and flood volume with useful skill.
The central value of CHA-Flood is not that it replaces process-based hydrological models, but that it organizes expert judgment, numerical simulation, and human review into an explicit, auditable, and reproducible workflow. By decomposing forecasting into multiple steps, in which LLMs are responsible for scenario interpretation and workflow orchestration while process-based models remain responsible for hydrological simulation, CHA-Flood preserves physical constraints that are often weakened in purely end-to-end deep-learning frameworks~\cite{kratzert2018rainfall, xiang2020rainfall}. It is also consistent with the forecasters-in-the-loop paradigm, in which human expertise remains essential for interpreting uncertain meteorological inputs and model outputs~\cite{tran2026value}. Transferring CHA-Flood to other basins or operational systems requires replacing or adapting not only the hydrometeorological records, hydrological model and parameter schemes, and warning thresholds to local conditions, but also the configuration of the six-dimensional similarity-retrieval method (including its variables, weights, and scoring criteria) and the local expert-derived reasoning rules. 
The main technical challenge is thus not only model portability, but also the systematic encoding of local expert knowledge into reusable and testable workflow components.

\subsection{Limitations}\label{sec:limitations}

We acknowledge some limitations in our study, the first concerning floods beyond the historical record. Step~1 derives its prior interval from retrieved analogs, so the interval is only as reliable as the support of the case library.
All five LLMs perform worst on Type~I floods (Figure~\ref{fig:LLMs_kfold_performance_costs}b), and the largest event in the record is underestimated by every LLM (Section~\ref{sec:PriorJudgment_validation}).
In a related stress test, standalone and hybrid hydrological models underestimated the largest observed flows when water years containing flows above a 5-year return-period threshold were withheld~\cite{acuna2025analyzing}.
For an event that exceeds all cases in the library, no retrieved analog can bound the prior interval. Although rule-based extrapolation is applied in this situation, a case-based method cannot be fully trusted to extrapolate beyond its library.
Because out-of-record floods matter most under a changing climate, we treat their robust handling as an open problem for CHA-Flood. One route is to extend CHA-Flood to additional basins and to pool analogs from hydrologically similar basins, so that the case library covers a wider range of extremes.
More broadly, difficulty in representing extremes is shared by data-driven Earth system models~\cite{watson2022machine, zhang2026physics}, whereas physics-constrained hybrid models generally behave better.
Therefore, future work should consider developing the workflow based on a hybrid modeling framework~\cite{reichstein2019deep}.
In such a framework, process-based models, machine learning models, Earth observation products, and LLM-based reasoning could be combined within the same operational workflow. In particular, incorporating a regional LSTM model, that is trained on a large number of catchments within a region, is especially promising, as previous studies have demonstrated its robust performance for streamflow prediction \cite{jia2026streamflow, kratzert2018rainfall}.
The LLM would not need to replace the hydrological model~\cite{tran2025ai}.
Instead, it could help determine when to rely on the process-based simulation, when to call a machine-learning correction model, and when the uncertainty is large enough to require additional expert review.
A more ambitious version of this framework may require hydrology-specific LLM adaptation or retraining, because general-purpose LLMs are not trained to represent hydrological process knowledge (like water balance), forecast uncertainty, and hydrological operational decision rules in a systematic way~\cite{kizilkaya2025towards}.
For example, Med-PaLM achieved expert-level clinical reasoning through medical instruction tuning of a general-purpose LLM~\cite{singhal2023large}, and GraphCast attained operational-level skill in global weather forecasting through domain-specific foundation-model training~\cite{lam2023learning}. These cases suggest that comparable gains could be achievable through hydrology-specific adaptation.

A further limitation of the current framework is that forecast uncertainty is still only partially quantified. Reasoning variability across the five LLMs provides one partial view of uncertainty in the language-model component. However, the current workflow still needs to examine the uncertainty from the LLM, especially the hallucination. One way is to examine the derivative of output to input to analyze the next-token gradient sensitivity~\cite{zhangnext}. Recent studies have raised similar concerns that data-driven models may learn predictive associations without capturing the underlying causal or physical mechanisms, and have proposed diagnostic strategies to distinguish whether models are learning ``why'' rather than only ``what''~\cite{shan2024assimilating}. Operational decisions depend not only on the predicted peak flow or flood volume, but also on how these quantities change under different rainfall, soil moisture, antecedent flow, or future climate conditions.
A model may achieve good predictive accuracy under historical conditions while still producing physically unrealistic responses when the forcing changes.
Therefore, future evaluations of CHA-Flood should include sensitivity tests to examine whether the forecast response is hydrologically reasonable.

Uncertainty from the meteorological variables like precipitation, model calibration, and the hydrological model structures are not well quantified. 
% Equifinality of hydrological parameters may downgrade the performance of the forecast~\cite{beven2001equifinality}. 
Meanwhile, the current simple hydrological model structure may also contribute a large portion of the uncertainty~\cite{clark2008framework}. Future work could design an idealized experiment to quantify the uncertainty from each component in CHA-Flood. This can be done by using an ensemble framework~\cite{cloke2009ensemble} of meteorological forcing variables, and different hydrological models. Machine-learning-based hydrological models like LSTM~\cite{kratzert2018rainfall}, transformer~\cite{yin2022rr}, or hybrid models~\cite{bhasme2022enhancing} can also be considered in CHA-Flood. % The current design of the workflow supports the transferablity from XAJ model to other hydrological models.

The current Step~2 design assumes that the hydrological model provides a finite library of pre-calibrated schemes. All calibration is completed offline in Step~0; Step~2 only selects among these schemes and does not recalibrate the model in real time. If a model is operated with one fixed parameter set, Step~0 can still calibrate and validate that set, and the model-independent Step~1 judgment can still provide prior intervals for peak flow and flood volume. The current Step~2 scheme selection is then omitted, whereas Step~3 can continue rolling forecasting by updating meteorological forcing and model states rather than parameters, and Step~4 can preserve review and audit records. The framework therefore loses the benefit of event-specific scheme selection but does not reduce to a prior-judgment check alone.
Data availability determines how far the workflow can proceed. When only archived observations are available, Step~0 can still construct the historical case library and calibrate the model, and completed events can be replayed to test the workflow retrospectively. Such replay is an offline evaluation rather than an operational forecast. For a current event, Step~1 can still estimate prior intervals for peak flow and flood volume if forecast rainfall characteristics and pre-event conditions are available. If a time-resolved rainfall forecast and initialized model states are also available, Step~2 can run the pre-calibrated schemes and complete scheme selection without using the future observed discharge. The Step~1 and Step~2 review gates therefore remain available under these conditions. What is lost without near-real-time rainfall and discharge telemetry is the observation-informed rolling update in Step~3 and its associated review gate. Step~4 can still document a static forecast or retrospective test, but it cannot report an observation-updated rolling forecast. CHA-Flood should therefore be regarded as an offline preparation and prior-judgment system where telemetry is unavailable, rather than as a complete real-time forecasting workflow.

\subsection{Way Forward}\label{sec:way_forward}

The next step is to move CHA-Flood from offline retrospective tests toward real operational use. Beyond these initial cross-basin tests, broader validation across dry, snow-dominated, regulated, and data-sparse basins is needed to establish the range over which the current skill framework remains reliable. Real-time testing during actual flood seasons is also essential, because operational value depends not only on retrospective accuracy but also on how forecasters interact with the system under time pressure, uncertain rainfall forecasts, incomplete observations, and institutional review procedures. Such deployment would allow CHA-Flood to accumulate forecaster feedback, update its case library, and refine local skills after each event. Looking further ahead, the standardization of hydrological skills, including version control, quality assessment, benchmark testing, and community sharing, may allow broader reuse of the framework and more systematic improvement of forecasting skills. CHA-Flood may also benefit from rapid progress in LLM-agent research, especially in memory and iterative skill refinement, which could improve long-term adaptation through continued interaction~\cite{cai2025building}. These capabilities could help CHA-Flood evolve from a static offline workflow into a continuously improving operational assistant, with possible future extensions to broader hydrological early warning~\cite{alfieri2013glofas}, reservoir operation~\cite{jain2023state, jia2019short}, and risk analysis~\cite{apel2009flood}.

To reduce uncertainty and improve forecast robustness, another important direction is to embed available Earth observation data into the CHA-Flood workflow. The purpose is to better constrain the uncertainty in meteorological forcing variables like precipitation~\cite{li2016application}, parameter tuning, and other components of the forecast workflow.
Satellite data provide unique information on hydrological states, such as soil moisture, that could help reduce forecast uncertainty, for example, through SMAP observations and SMAP-derived surface and root-zone soil moisture products~\cite{entekhabi2010smap,reichle2017smap}.
Such information could be used to quantitatively update the event description, revise the prior judgment, or constrain the rolling forecast~\cite{jadidoleslam2021data}.
For example, after a new flood event occurs, the corresponding observations could be added to the case library. Comparisons between new observations and forecasts could also be used to refine event-retrieval rules and to support Step~3 updating.
Such assimilation could also help test the robustness of LLM-supported reasoning by comparing workflow outputs with satellite observations.
However, this also requires careful treatment of observation error and spatial representativeness, because the value of additional observations depends on whether they are consistent with the model states and forecast variables.

LLM agents for hydrology remain an emerging research direction. CHA-Flood provides one step toward this direction by showing how language models can be embedded as workflow orchestrators that interact with hydrological models, case libraries, expert rules, and human forecasters. However, further exploration is needed to evaluate the reliability, failure modes, uncertainty, and transferability under diverse hydrological and operational conditions. This may require transparent skill designs, community benchmarks, uncertainty-aware evaluation, and close collaboration between hydrologists, forecasters, and AI researchers. These efforts may evaluate whether LLM agents can become trustworthy components of next-generation hydrological forecasting and decision-support systems.

\section{Conclusions}\label{sec:conclusion}

In this study, we proposed CHA-Flood, a skill-orchestrated agent framework that embeds LLMs into a physically based flood forecasting workflow translating the tacit expertise of operational forecasters into an auditable and reproducible procedure.
CHA-Flood was tested on the South Yamhill River basin (Oregon, United States) using 113 historical flood events (1995--2019) for scheme preparation and 14 recent events (2020--2024) for validation. The Step 1 scenario judgment skill captured observed peak flow and flood volume within a 5\% tolerance margin in 10 and 11 out of 14 validation events, respectively, and stratified 5-fold cross-validation over all 129 historical events yielded Pearson correlations of 0.62 for peak flow and 0.84 for flood volume. The Step 2 scheme selection, guided by these prior ranges, improved KGE by 0.023--0.154 for events where the scheme was updated, with simulated peak flow and flood volume falling within the prior judgment ranges for 14 and 13 out of 14 events.
All five state-of-the-art LLMs successfully executed the full workflow with comparable judgment accuracy, though moderate performance variation and substantial cost differences emerged across models. 
Removing the language-model narrowing with the rest of the CHA-Flood workflow unchanged lowered median KGE from 0.723--0.753 to 0.662 for every LLM, indicating that the improvement came from LLM reasoning rather than from the choice of LLM.
Ten repeated runs with each of the five LLMs showed similar central skill across models, with tightly clustered prior ranges for GPT-5.4 and larger run-to-run variation for the other models at a few class-boundary events.
A notable limitation arose for Type V floods with weak hydrological signals, where Step 1 misjudgment propagated into Step 2 and degraded the forecast hydrograph, highlighting that prior-judgment accuracy is a critical prerequisite for the overall forecasting chain.
Supplementary tests in two hydrologically contrasting basins confirmed that the shared workflow and data contracts were technically transferable, while their predictive performance emphasized the continued need for basin-specific knowledge, retrieval rules, and hydrological model adaptation.

These results suggest that CHA-Flood's primary contribution lies not in direct numerical prediction, but in the transparent orchestration of expert judgment, physically based simulation, and human review within an auditable workflow.
Rather than replacing hydrological models, CHA-Flood formalizes tacit forecaster expertise into operational skills with explicit boundaries that constrain language model reasoning to complement physical simulation, offering a viable paradigm for next-generation intelligent flood forecasting.
Nonetheless, CHA-Flood remains a first step toward a fully operational system.
Extending the evaluation beyond these three basins, integrating hybrid modeling with Earth observation data, and enabling continuous case-library updating through cross-basin retrieval could further strengthen its physical consistency and operational adaptability.

%%%%%%%%%%%%%%%%%%%%%%%%%%%%%%%%%%%%%%%%%%%%%%%
%  BACK MATTER
%%%%%%%%%%%%%%%%%%%%%%%%%%%%%%%%%%%%%%%%%%%%%%%

\section*{Open Research Statement}
The hourly hydrometeorological records used in this study were obtained from the CAMELSH data set~\cite{tran2025camelshdata}, available at https://zenodo.org/records/16763144. The primary analysis used USGS gauge 14194150 (South Yamhill River at McMinnville, Oregon, United States), and the supplementary cross-basin experiments used gauges 11532500 (Smith River near Crescent City, California, United States) and 08189500 (Mission River at Refugio, Texas, United States). Streamflow observations for all three gauges are publicly available through the USGS National Water Information System (https://waterdata.usgs.gov/).
Processed event data, historical case libraries, calibrated XAJ parameters and benchmark outputs are archived on Zenodo~\cite{shan_2026_23723094} (https://doi.org/10.5281/zenodo.20644504). The archive also includes validation prompts, figure-reproduction scripts and a compiled program for replaying the deterministic rules in Steps~1 and~2. These materials support auditing the reported results and reproducing figures and tables. 

\section*{Conflict of Interest Declaration}
The authors declare there are no conflicts of interest for this manuscript.

\newpage
\acknowledgments
\indent We gratefully acknowledge the Editor, Prof. Marc Bierkens, and the three anonymous reviewers for their careful reading of our manuscript and their constructive comments.
We also thank Tao Wang (Department of Hydraulic Engineering, Tsinghua University) for insightful discussions on formalizing forecaster expertise in CHA-Flood.

We thank the Hydro90 community for bringing together this group of authors and for enabling the rapid progress of this work.
This work received in-kind computational support in the form of token credits from Zeta Frontier Technology (Hangzhou) Co., Ltd. The sponsor had no role in study design, methodology, analysis, interpretation of results, manuscript writing, or the decision to submit the manuscript.

%%%%%%%%%%%%%%%%%%%%%%%%%%%%%%%%%%%%%%%%%%%%%%%
% REFERENCES
%%%%%%%%%%%%%%%%%%%%%%%%%%%%%%%%%%%%%%%%%%%%%%%
%
% Please use ONLY~\cite and~\citeA for reference citations.
%~\cite for parenthetical references
%~\citeA for in-text citations
% DO NOT use other cite commands (e.g.,~\citet,~\citep,~\citeyear, \nocite,~\citealp, etc.).

\newpage
\bibliography{refs}

\section*{References From the Supporting Information}

\noindent Gong, J., Liu, X., Yao, C., Li, Z., Weerts, A. H., Li, Q., Bastola, S., Huang, Y., \& Xu, J. (2025). State updating of the Xin'anjiang model: Joint assimilating streamflow and multi-source soil moisture data via the asynchronous ensemble Kalman filter with enhanced error models. \textit{Hydrology and Earth System Sciences}, \textit{29}, 335--360. \url{https://doi.org/10.5194/hess-29-335-2025}

\noindent Luo, Y., Zhou, Y., Xu, H., Chen, H., Chang, F.-J., \& Xu, C.-Y. (2024). Enhancing physically-based flood forecasts through fusion of long short-term memory neural network with unscented Kalman filter. \textit{Journal of Hydrology}, \textit{641}, 131819. \url{https://doi.org/10.1016/j.jhydrol.2024.131819}

\noindent Wang, Y., Liu, W., Zhu, F., Xu, B., Li, W., Tong, J., \& Zhong, P.-a. (2025). Modeling the cumulative impact of minor impoundments on watershed hydrology: Enhancements to the Xin'anjiang model. \textit{Journal of Hydrology}, \textit{660}, 133399. \url{https://doi.org/10.1016/j.jhydrol.2025.133399}

%%%%%%%%%%%%%%%%%%%%%%%%%%%%%%%%%%%%%%%%%%%%%%%
%% Optional Appendices
%%%%%%%%%%%%%%%%%%%%%%%%%%%%%%%%%%%%%%%%%%%%%%%

%\appendix
%\section{Appendix 1}
%Appendix text.

\end{document}